# Superconductivity for Magnets

*R. Flükiger[1]*
CERN, Geneva, Switzerland

**Abstract**
The present state of development of a series of industrial superconductors is reviewed in consideration of their future applications in high field accelerator magnets, with particular attention on the material aspect. The discussion is centred on $Nb_3Sn$ and $MgB_2$, which are industrially available in a round wire configuration in kilometre lengths and are already envisaged for use in the LHC Upgrade (HL-LHC). The two systems Bi-2212 and R.E.-123 may be used in magnets with even higher fields in future accelerators: they are briefly described.

## 1 Introduction

Superconducting high field magnets are based on wires or tapes consisting of a variety of materials. In this manuscript, selected current-carrying properties of various commonly used superconducting wires for accelerators are described. The main subjects will be the two systems $Nb_3Sn$ and $MgB_2$; round wires in these materials can already be produced industrially in long lengths. A brief presentation of the two other known systems, Bi-2212 and YBaCuO, will follow. Bi-2223 tapes are also available in kilometre lengths, but are not envisaged for use in future accelerators and are not included here.

## 2 Multifilamentary $Nb_3Sn$ wires: preparation techniques

$Nb_3Sn$ is in the cubic phase, with $T_c$ = 18.2 K, and belongs to the category of Low-Temperature Superconductors (LTS). Its upper critical field $B_{c2}$ varies between 25 T (measured in binary $Nb_3Sn$) and 30 T (after alloying with Ti or Ta). Industrial $Nb_3Sn$ wires always contain a certain content of additives (~1.5 at.% Ti or ~3 at.% Ta) in order to achieve a maximum value of $B_{c2}$.

$Nb_3Sn$ multifilamentary wires can be prepared using several techniques. The most common ones are (a) the 'bronze route' (the reaction of Cu-Sn bronze and Nb), (b) the Powder-In-Tube (PIT) technique (where $Nb_3Sn$ is formed by a reaction between Nb and $NbSn_2$), and (c) the RRP technique (a direct reaction between Nb and Sn, based on the internal Sn diffusion mechanism). In contrast to the bronze route, the two other techniques, PIT and RRP, do not require intermediate heat treatments, as represented schematically in Fig. 1.

---

[1] rene.flukiger@cern.ch

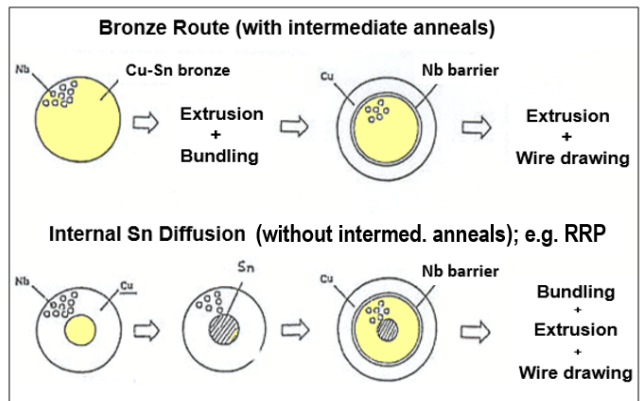

**Fig. 1:** Schematic representation of the processing for bronze route and RRP Nb$_3$Sn wires

Out of the three processing methods followed for the fabrication of Nb$_3$Sn wires, only the two with the highest $T_c$ values have been retained at CERN for further studies in consideration of the LHC Upgrade (or HL-LHC) quadrupoles: the RRP wires from Oxford Instruments (USA) and the PIT wires from Bruker (Hanau, Germany) (see Figs. 2 and 3). The investigations at CERN involved wires with diameters between 0.7 and 1.0 mm, with Cu/non-Cu ratios ranging from 1.15 to 1.25, with unit lengths reaching 800 m. The RRR ratio, important for the thermal stability of the wire, was 150 in both types of wire. The filament diameters were between 40 and 60 mm for the RRP wires and between 35 and 50 mm for the PIT wires.

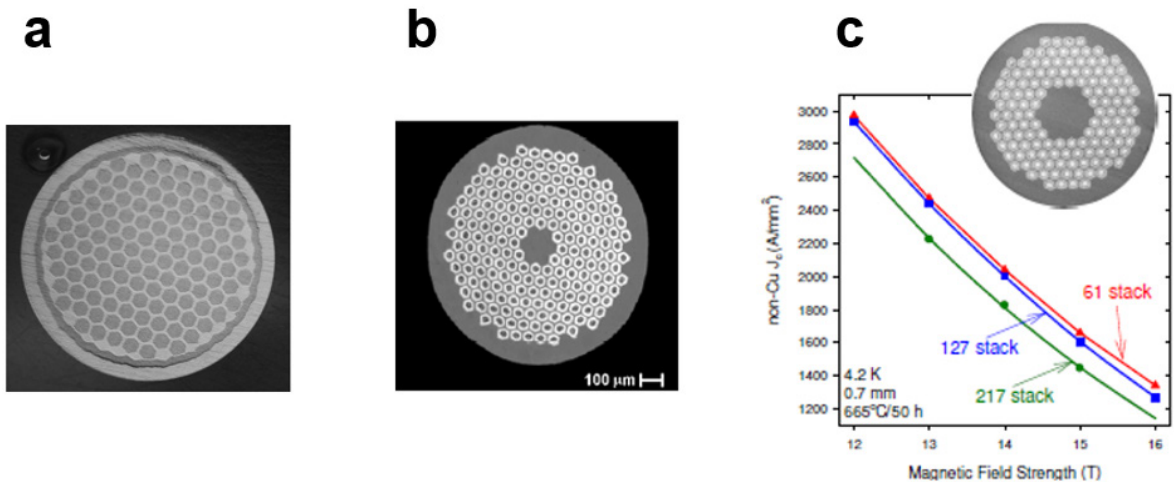

**Fig. 2:** Cross-sections of (a) bronze route, (b) PIT, and (c) RRP multifilamentary Nb$_3$Sn wires. The non-Cu critical current densities up to 16 T are shown in (c) for RRP wires. (Courtesy of J.M. Parrel, OST.)

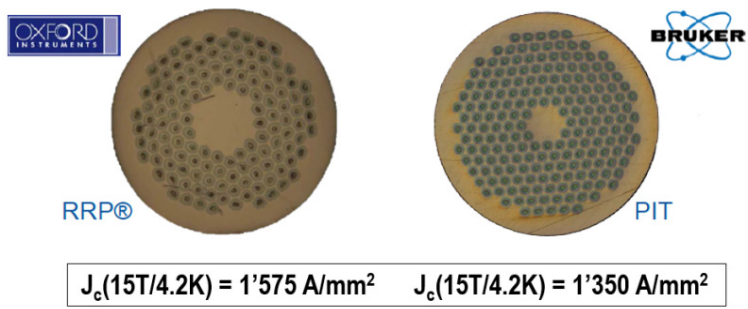

**Fig. 3:** The two Nb$_3$Sn wire types tested at CERN in view of LH-LHC quadrupoles: RRP wires from Oxford Instruments (USA) and PIT wires from Bruker (Hanau, Germany).

## 2.1 Uniaxial tensile stress

In view of the use of $Nb_3Sn$ wires in large magnets with very high currents and fields, the question arises about the effect of the strong Lorentz forces under these extreme conditions. The 3D stress situation in a wire has been treated by analysing two cases: (a) the uniaxial tensile stress and (b) the compressive stress. Although $Nb_3Sn$ has the highest electronic density of states, $N(E_F)$, among all A15-type compounds, it is thought that the variation of this property is of minor importance in understanding the effects of Lorentz forces. Indeed, it has been shown [1] that the effect of uniaxial applied stress on $J_c$ is mainly correlated to changes in the phonon spectrum. It was found that the main variation of $J_c$ under uniaxial tensile stress is primarily sensitive to non-hydrostatical stress components, the hydrostatical components having a markedly lower influence.

It has been shown by neutron diffraction measurements [2] that, during the cooling process, the A15 lattice undergoes an increasing degree of elastic distortion into a tetragonal phase: the distortion starts at ~500 K and is maximum at 4.2 K, where $(1 - c/a) \sim 0.2\%$. When applying a stress in the axial direction, it was thus argued that the precompression would gradually diminish, to reach a 'stress-free' state at a given stress value. This was confirmed by a direct measurement of the crystallographic changes of the A15 phase inside the filaments when applying tensile stresses at 10 K, first performed by means of X-ray diffraction, on a flat, monofilamentary bronze route $Nb_3Sn$ wire [2]. The bronze:Nb ratio of this sample was 7:1, which explains the relatively high value of $\varepsilon_m = 0.5\%$ (Fig. 4), at which the measured value of $I_c$ reaches a maximum (also shown in Fig. 4). This particular design was chosen for the direct observation of the effect by diffractometry. The same authors showed later by neutron diffraction that this behaviour is also representative for multifilamentary $Nb_3Sn$ wires.

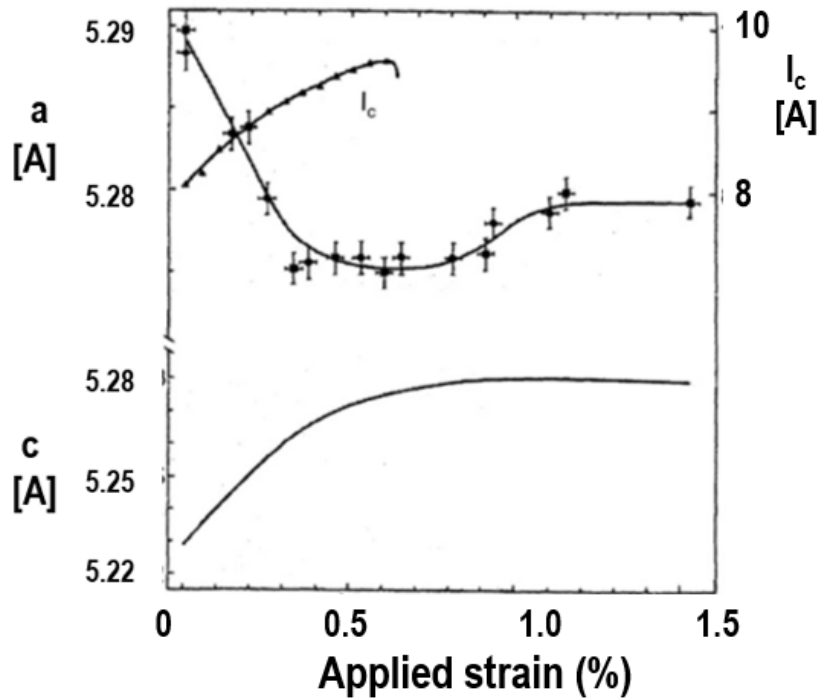

**Fig. 4:** Critical current $I_c$ and lattice parameters $a$ and $c$ on a monofilamentary flat $Nb_3Sn$ wire vs. the uniaxial applied strain $\varepsilon$. The variation of $a$ is directly measured, and $c$ is calculated after correction for the volume compression $\Delta V(\varepsilon)$: $\Delta V(0) = 0.6\%$ and $\Delta V(1.0\%) = 0.8\%$ [2].

Various measuring devices have been developed to measure $J_c(\varepsilon)$, the variation of $J_c$ as a function of the applied strain $\varepsilon$: the linear strain rig [3], the Pacman (University of Twente), and the modified Walters spiral, which was developed at the University of Geneva [4] and later also at NIST, Boulder, CO, USA. The Walters spirals for the measurement of $J_c(B,\varepsilon)$ on wires and tapes are shown in Fig. 5.

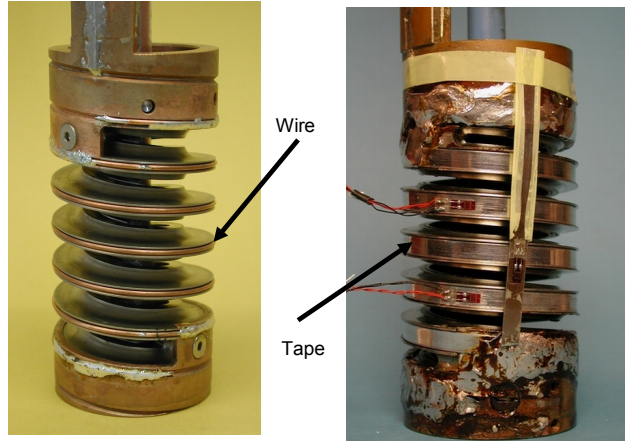

**Fig. 5:** Modified Walters spirals developed at the University of Geneva to measure the effect of uniaxial tensile strain on round Nb$_3$Sn wires (left) and R.E.-123-coated conductor tapes (right) up to 21 T at 4.2 K. The characteristics of both Walters spirals are $I_c \leq 1000$ A, with wire lengths $\leq 0.8$ m, thus allowing a strong $I_c$ criterion: 0.1 or 0.01 $\mu$V·cm$^{-1}$ [4].

The uniaxial force on the wire or tape is obtained by a rotation of the central axis of the spiral. The variation of $J_c(\varepsilon)$ can be understood by considering that the Nb$_3$Sn filaments in the wire are under precompression, as a result of the differential thermal contraction between the Nb$_3$Sn phase and the surrounding Cu or Cu-Sn bronze. Since the linear thermal contraction value $\alpha$ of Nb$_3$Sn is $8 \times 10^{-6}$ K$^{-1}$, i.e. considerably lower than the value for Cu or Cu-Sn ($18 \times 10^{-6}$ K$^{-1}$), the cooling from ~1000 K (the reaction temperature) to 4.2 K (the operation temperature) results in a compression of the matrix on the filaments (precompression). The resulting reversible distortion has a strong effect on the physical properties, in particular on the critical current density values $J_c$. For a given temperature, the latter can now be described as depending on two parameters: $J_c(B,\varepsilon)$ (see Fig. 6).

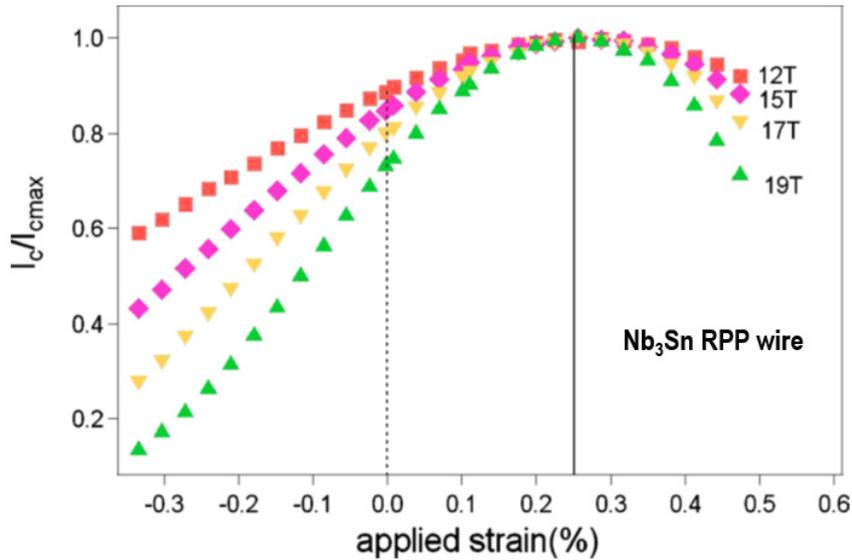

**Fig. 6:** Normalized $I_c$ vs. applied strain for a RRP Nb$_3$Sn wire at $T = 4.2$ K and for different fields. $I_c$ behaves reversibly in the shown strain window ($\varepsilon_m = 0.25\%$). The critical currents at zero applied strain (0.1 $\mu$V·cm$^{-1}$) are marked by the dashed line: 278.4 A (12 T), 156.6 A (15 T), 95.9 A (17 T), 50.4 A (19 T). (From [5].)

The question about the possible effect of uniaxial tensile stresses on flux pinning can be answered by representing the normalized pinning force $F_p/F_{p,max}$ as a function of the reduced magnetic field $b = B/B_{c2}$; as shown [5] for a RRP wire, all data fall on a universal curve, reflecting that the pinning mechanism is not influenced by the application of uniaxial tensile strain (see Fig. 7). This confirms that the effects of uniaxial stress on $J_c$ are essentially due to the change of $B_{c2}$.

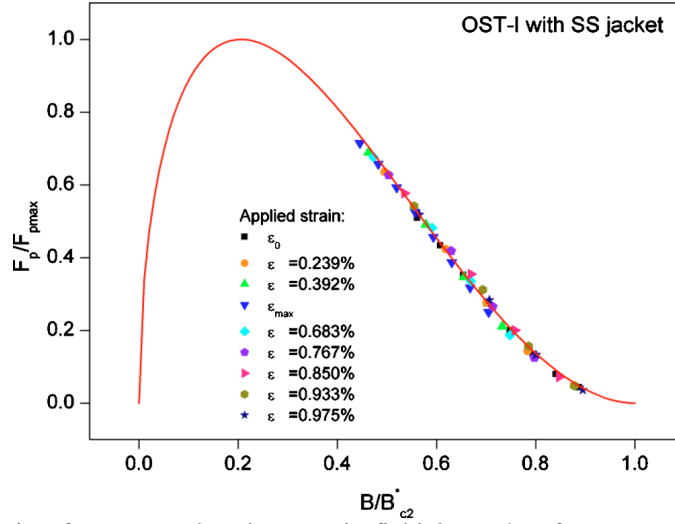

**Fig. 7:** Normalized pinning force vs. reduced magnetic field $b = B/B_{c2}$ for a RRP Nb$_3$Sn wire from Oxford Instruments (OST). All data fall on a universal curve, indicating that the pinning mechanism is not influenced by the application of uniaxial tensile strain. (From [6].)

It can be shown that a stronger criterion for $I_c$ (defined by $\mu$V·cm$^{-1}$) leads to an earlier detection of nanocracks: the longer wire length in the Walters spiral allows a deeper insight, as shown by Fig. 8, on a bronze route Nb$_3$Sn wire of 0.8 mm diameter produced by Furukawa for ITER [6]. It follows that the early detection of the irreversible strain limit $\varepsilon_{irr}$ (the beginning of the nanocrack formation) depends on the $I_c$ criterion. It follows from Fig. 8 that $I_c$ does not depend on the strain criterion up to 0.6–0.7%. However, the effects of the various criteria are seen after releasing the strain $\varepsilon$, as shown in Fig. 8.

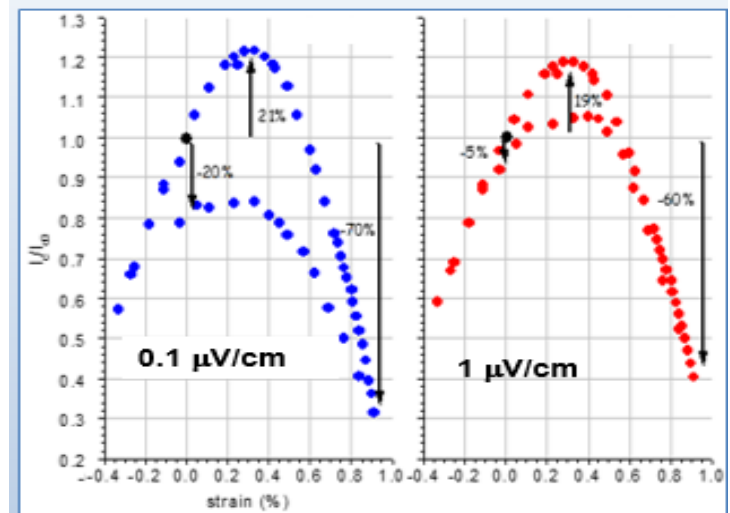

**Fig. 8:** Variation of $I_c/I_{co}$ vs. uniaxial strain $\varepsilon$ at 4.2 K and 13 T for a Furukawa bronze route wire, showing the earlier detection of the irreversible strain limit due to nanocracks for the stronger $I_c$ criterion. (From [6].)

A correlation between different parameters, e.g. temperature, field and strain dependence superconducting parameters, can be found by a parametrization proposed by L. Bottura (CERN), starting with a formula due to Summers:

$$J_c(B,T,\varepsilon) = \frac{C_{Nb_3Sn}(\varepsilon)}{\sqrt{B}}\left[1 - \frac{B}{B_{c2}(T,\varepsilon)}\right]^2 \left[1 - \left(\frac{T}{T_{c0}(\varepsilon)}\right)^2\right]^2,$$

$$\frac{B_{c2}(T,\varepsilon)}{B_{c20}} = \left[1-\left(\frac{T}{T_{c0}(\varepsilon)}\right)^2\right]\left\{1-0.31\left(\frac{T}{T_{c0}(\varepsilon)}\right)^2\left[1-1.77\ln\left(\frac{T}{T_{c0}(\varepsilon)}\right)\right]\right\},$$

$$C_{Nb_3Sn}(\varepsilon) = C_{Nb_3Sn,0}\left(1-\alpha_{Nb_3Sn}|\varepsilon|^{1.7}\right)^{1/2},$$

$$B_{c20}(\varepsilon) = B_{c20m}\left(1-\alpha_{Nb_3Sn}|\varepsilon|^{1.7}\right),$$

$$T_{c0}(\varepsilon) = T_{c0m}\left(1-\alpha_{Nb_3Sn}|\varepsilon|^{1.7}\right)^{1/3},$$

where $\alpha_{Nb_3Sn}$ is 900 for $\varepsilon = -0.003$, $T_{c0m}$ is 18 K, $B_{cm0}$ is 24 T, and $C_{Nb_3Sn,0}$ is a fitting parameter equal to 60 800 AT$^{1/2}$·mm$^{-2}$ for a value of $J_c = 3000$ A·mm$^{-2}$ at 4.2 K and 12 T.

### 2.2 Effect of transverse stresses

The effect of transverse stress on the superconducting properties, and in particular on the current-carrying capability, is much stronger than that of uniaxial stress. This is illustrated in Fig. 9, which shows the dependence of the ratio $I_c/I_{c0}$ of a bronze route wire as a function of the applied stress, where $\sigma_a$ is the uniaxial stress and $\sigma_t$ is the transverse stress. It is clearly seen that a maximum of $I_c/I_{c0}$ is reached for transverse stresses $\sigma_t$ of the order of 20 MPa, in contrast to uniaxial stresses, where $\sigma_a$ is observed at 150 MPa; thus, $\sigma_t(I_{cm}) \ll \sigma_a(I_{cm})$ [7].

A particularly important problem that arises when considering the use of Nb$_3$Sn wires in accelerators concerns their behaviour under the effect of transverse stress produced by the large Lorentz forces. This problem has been studied at the University of Geneva, within the framework of the Ph.D. work [8] by G. Mondonico, who studied the variation of $J_c$ in three load cases. These cases are:

1) a wire pressed between two parallel walls (this is the worst possible case and is never encountered);
2) a wire simultaneously pressed by four walls without epoxy;
3) same as case 2), but with epoxy (thus simulating the stress on a wire in a Rutherford cable).

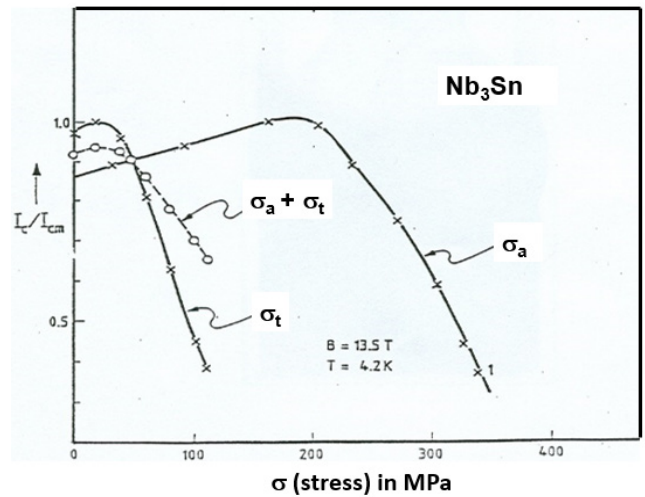

**Fig. 9:** Comparison between uniaxial tensile stress and compressive stress on the same bronze route wire, showing a considerably stronger effect due to the compressive stresses [7].

The result of this investigation is represented in Fig. 10; it shows that the epoxy has a very beneficial effect, rendering the stress distribution more hydrostatic, thus reducing the transverse stress acting directly on the wires in the Rutherford cable.

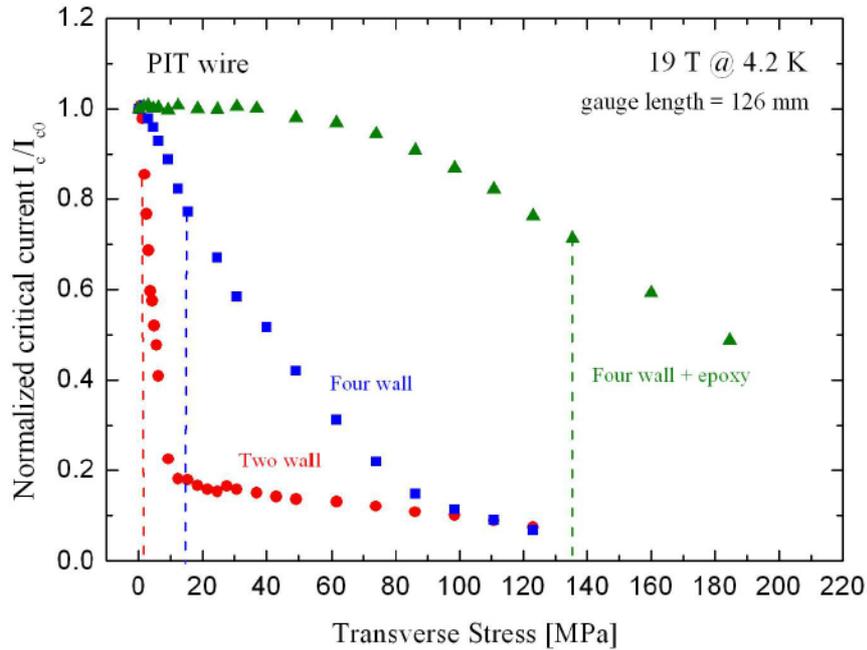

**Fig. 10:** $I_c/I_{c0}$ vs. $\sigma_t$ at 19 T and 4.2 K for the PIT wire in the three different compressive stress configurations. The irreversible limits are indicated by dashed lines. (From [8].)

### 2.3  Conclusions about Nb$_3$Sn wires

As the present contribution is in the form of a lecture, it contains a limited choice of subjects and references, just necessary to present the main physical properties of Nb$_3$Sn wires. The main features of the behaviour of Nb$_3$Sn wires can be summarized as follows.

- Small solenoids for NMR based on Nb$_3$Sn can reach 23.5 T (1 GHz). In large dipoles or quadrupoles (e.g. for LHC Upgrade), the achievable field is limited to 13–15 T, due to the limited available space, large Lorentz forces, and thermal stability requirements.
- The amount of Nb$_3$Sn wire in a magnet increases strongly with the produced field: at 20 T, it is five times more than for 12 T.
- The critical current density at small fields is dictated by flux pinning due to A15 grain boundaries (size: 50–100 nm). At high fields, $J_c$ is mainly influenced by the value of $B_{c2}$.
- Both internal Sn (RRP) and Powder-in-Tube (PIT) wires satisfy or almost satisfy the conditions for LHC Upgrade accelerator magnets: $J_c$ = 1500 A·mm$^{-2}$ at 4.2 K/15 T. Bronze route wires do not reach these high $J_c$ values, but are best suited for the 'persistent mode' operation of NMR magnets.

## 3  The MgB$_2$ system

The MgB$_2$ phase ($T_c$ = 39 K) crystallizes into a hexagonal lattice (Fig. 9), with $a$ = 0.30834 nm and $c$ = 0.35213 nm. This compound has been chosen for the high current connection between the surface and the dipoles and quadrupoles of the LHC Upgrade. These connections will carry a current of 13 000 A, at an operation temperature of 10 K. The phase diagram of the Mg-B system (in Fig. 11) shows the presence of a gas phase, which has to be taken into account when preparing MgB$_2$ wires.

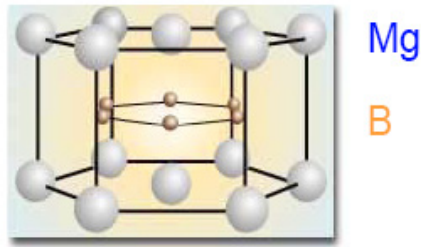

**Fig. 11:** The hexagonal structure of the MgB$_2$ phase

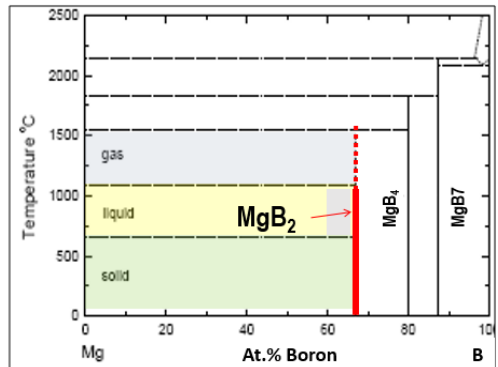

**Fig. 12:** The Mg-B high-temperature phase diagram, showing a gaseous domain above 1100°C

The superconducting properties of the compound MgB$_2$ are anisotropic, similar to those of High-Temperature Superconductor (HTS) compounds. However, the anisotropy of MgB$_2$ is much smaller, and its behaviour is more like that of Low-Temperature Superconductor (LTS) compounds. MgB$_2$ has some peculiar properties; in particular, it exhibits two superconducting gaps. This important fact has been proven by a series of physical measurements, e.g. by low-temperature calorimetry, and has been theoretically described. For the description of the current-carrying capability, however, the larger gap is the more important one, and most superconducting properties can be described by the BCS model, where the electron–phonon interaction leads to Cooper pairs.

### 3.1 The irreversibility field, $B_{irr}$

An 'irreversibility field' $B_{irr}$ can be found in high-temperature superconductors which is markedly different from the value of the upper critical field $B_{c2}$. This is not the case for Nb$_3$Sn, where $B_{irr}$ and $B_{c2}$ almost fall together. The determination of $B_{irr}$ can be based on the measurements of the electrical resistivity $R(T)$. The definition is somewhat arbitrary: usually, the values are taken at 10% and 90%, or at the normal-state resistivity value $R_n$, for $B_{c2}$ and $B_{irr}$, respectively. The dependence of $B_{c2}$ and $B_{irr}$ on the temperature for binary *ex situ* MgB$_2$ wires is illustrated in Fig. 13.

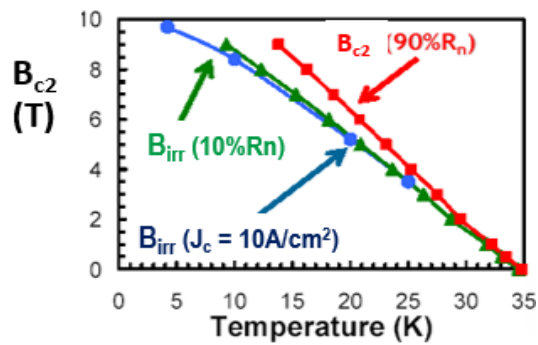

**Fig. 13:** Dependence of $B_{irr}(T)$ and $B_{c2}(T)$ in a binary MgB$_2$ wire (field // to *ab* plane)

## 3.2 The anisotropy of $B_{c2}$ in $MgB_2$

The physical properties of $MgB_2$ show a marked anisotropy. As can be seen from Fig. 14 for single crystals, the value of $B_{c2}^{\perp}$ perpendicular to the *ab* planes is markedly lower than $B_{c2}^{//}$, the value parallel to *ab*. In carbon alloyed $MgB_2$, the anisotropy is lowered. At sufficiently high carbon contents (too high for the optimization of the critical current density) it disappears.

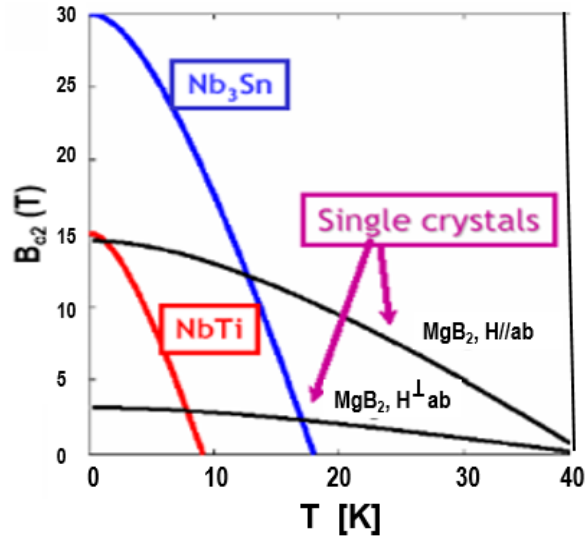

**Fig. 14:** $B_{c2}(T)$ for Nb−Ti, $Nb_3Sn$, and alloyed $MgB_2$ single crystals. The latter show a marked anisotropy with respect to the applied field.

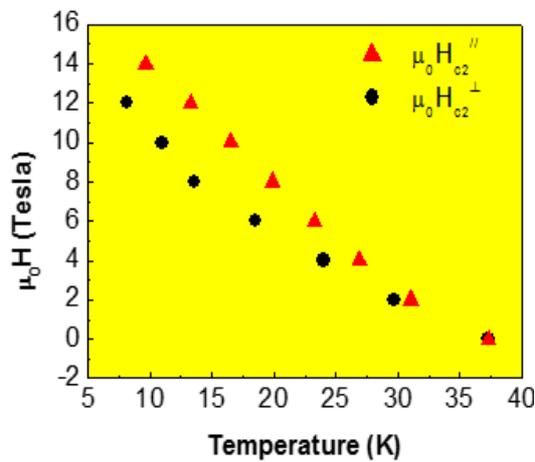

**Fig. 15:** Anisotropy of the upper critical field $B_{c2}$ of binary $MgB_2$ wires

## 3.3 Effect of alloying on $J_c$

A strong enhancement of $J_c$ was observed after adding SiC nanopowders to the initial Mg+B powder mixtures [9]. The effect of SiC additives on the transport $J_c$ of $MgB_2$ wires at 4.2 K is shown in Fig. 16: $J_c$ increases at high fields, but decreases at low fields. The effect of additives on $J_c$ of $MgB_2$ has been the subject of numerous investigations [10]. Single-crystal observations show that the substitution of C on the B lattice sites leads to an enhanced residual resistivity $\rho_0$, which is correlated to the observed enhancement of the upper critical field, $B_{c2}$, and the irreversibility field, $B_{irr}$, in $MgB_2$. The enhancement of $\rho_0$ with lattice disorder in alloyed $MgB_2$ wires was clearly recognized as the dominant effect enhancing the critical current density, $J_c$, at high magnetic fields. The vortex pinning behaviour of $MgB_2$ is only slightly affected by the presence of C on the B sites. This behaviour is similar to that encountered in $Nb_3Sn$: alloying by Ta and Ti leads to a reduction of the electronic mean free path and thus to an enhancement of $B_{c2}$, with the consequence of an enhancement of $J_c$ at high fields. At lower fields, however, $J_c$ is always lower in alloyed wires [11]. The reasons for the decrease of pinning force at low fields are not explicitly known. However, there are reasons to think that the grain boundaries are affected by the presence of additives, in analogy to the case of Nb$_3$Sn wires. There is a direct consequence of the behaviour illustrated in Fig. 16 in view of the LINK application in the LHC Upgrade: since the MgB$_2$ wires in this application at 10 K will be exposed to very low fields only, binary rather than ternary alloyed MgB$_2$ wires constitute the logical choice. In this particular case, the preference will be given to *ex situ* wires [12], which in addition are also mechanically the most robust ones.

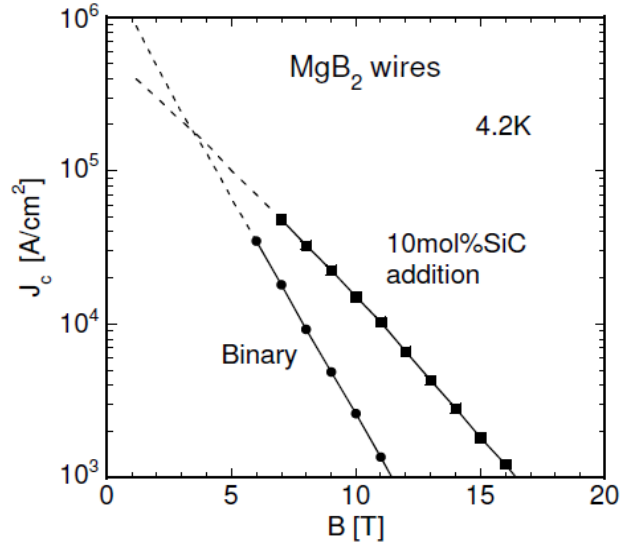

**Fig. 16:** Comparison between the variation of $J_c$ vs. the applied field B for binary and SiC added *in situ* MgB$_2$ wires [11].

### 3.4 Fabrication of MgB$_2$ wires

Three main routes for the fabrication of MgB$_2$ wires are established, based on the Powder-in-Tube (PIT) method: (a) the *ex situ* method [13] (Columbus, Genoa, Italy), (b) the *in situ* method [14] (Hypertech, Columbus, Ohio, USA), and (c) the Internal Mg Diffusion process (IMD) [15, 16]. Wires of several kilometre lengths have been fabricated by both *ex situ* and *in situ* processing; IMD wires are still being optimized. The fabrication steps are shown schematically in Fig. 17, which emphasizes the fact that the main difference between *ex situ*, *in situ*, and IMD processing resides in the initial powder mixture introduced to the metallic tubes. The tubes may consist of Fe, Ni or Monel, with Nb as a barrier material.

Recently, it was found that tubes consisting of pure Ti do not require a barrier, which is an advantage for the fabrication procedure.

In *ex situ* wires, the powder mixture consists of previously reacted MgB$_2$ powder particles annealed at 900°C at the end of the deformation process [13]. The *in situ* process starts with a mixture of Mg and B powders, the reaction at 650°C occurring at the end of the deformation process [14]. In the IMD process, first introduced by G. Giunchi [16], Mg rods are introduced inside a 'tube' consisting of pressed boron powder, the reaction occurring after deformation. The main processing steps of bundling and deformation to the wire are similar for the three methods. There is an important difference between the mass density inside the MgB$_2$ filaments obtained by these three methods: in IMD wires the mass density is highest and is very close to 100%. The mass density in *ex situ* and *in situ* wires is considerably smaller: ~75% and ~50%, respectively. It will be shown in the following paragraph that in MgB$_2$ wires—prepared by powder metallurgical processing—the mass density is correlated to $J_c$.

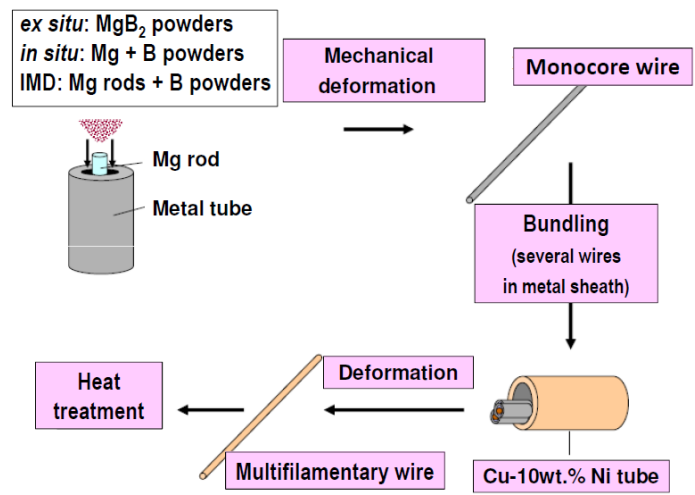

**Fig. 17:** The three different methods of MgB$_2$ wire fabrication by powder metallurgical processing

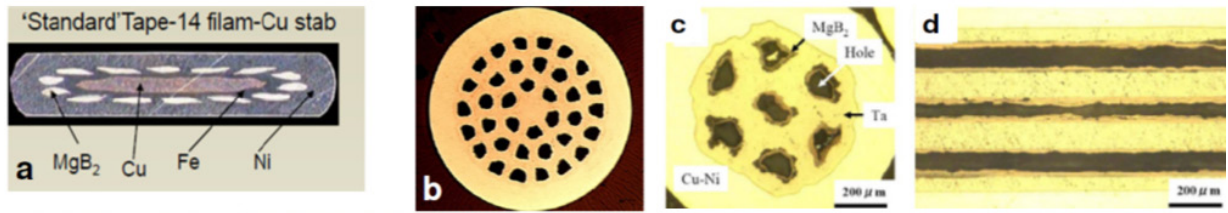

**Fig. 18:** Typical cross-sections of MgB$_2$ multifilamentary wires prepared by three different industrial processes: (a) *ex situ*, (b) *in situ*, and (c), (d) IMD processing [11].

*Ex situ* wires can also be prepared in a round shape. Various MgB$_2$ wire configurations tested at CERN for the LINK project in LHC Upgrade are shown in Fig. 17 (from the presentation by A. Ballarino at this CAS).

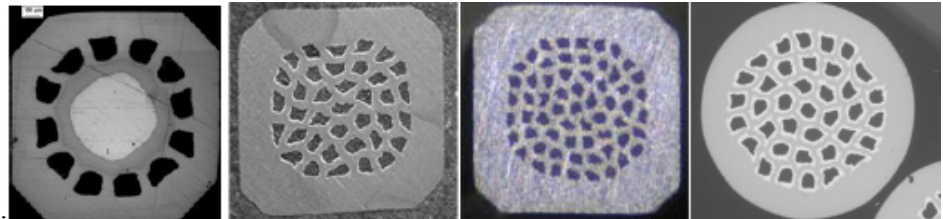

**Fig. 19:** Various cross-sections of round, Cu-stabilized *ex situ* MgB$_2$ wires produced by Columbus for the LINK project at CERN [17].

## 3.5 Effect of cold pressing on $J_c$ of MgB$_2$ wires

Wires with powder metallurgically prepared filaments contain a certain amount (depending on the fabrication path) of voids. As mentioned in Section 3.4, IMD wires present almost no voids, but in *ex situ* and *in situ* wires they are of the order 25 and 50 vol.%, respectively. Attempts to increase the mass density by performing the final heat treatment under high Ar pressure (0.2 MPa) or in hard metal anvils (>10 GPa) yielded a moderate increase of $J_c$ only: these methods are not applicable for industrial purposes because the gain in $J_c$ is too small to cover the costs of the pressure application. An alternative to these processes consists in submitting the wires to a high pressure at room temperature after the final deformation, just before the high-temperature treatment. Cold high-pressure processing [18], consisting of a simultaneous application of high pressure on four sides of square wires, has led to an enhancement of mass density (up to 90%) and thus of $J_c$ for *in situ* MgB$_2$ wires [19], as a result of better connectivity. As shown on the left of Fig. 20, an enhancement of $J_c$ at 4.2 K by a factor 2 at all fields was obtained after cold pressing at 1.48 GPa; the same $J_c$ value can be obtained at fields 2 T higher after cold pressing. On the right of Fig. 20, considerably higher enhancements of $J_c$ at 20 and 25 K are obtained on cold pressed wires, by factors 5 and 8, respectively.

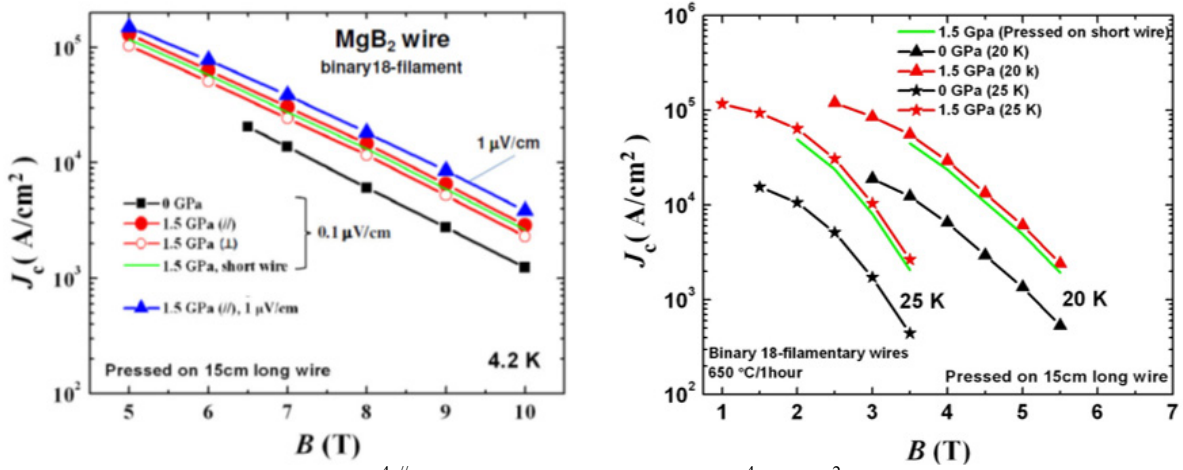

**Fig. 20:** Variation of the fields $B(10^4)^{//}$ (field values where $J_c = 1 \times 10^4$ A·cm$^{-2}$) at 4.2, 20, and 25 K) for $C_4H_6O_5$ alloyed MgB$_2$ tapes (Monel sheath, Nb barrier) after cold densification at various pressures. Reaction conditions: 600°C/4 h (600°C/4 h) [19].

### 3.6 Present state of $J_c$ vs. $B$ in MgB$_2$ wires

The highest values of $J_c$ known so far have been obtained by Li *et al.* [20], who used a modified IMD method, yielding almost 100% dense filaments. Their values are shown in Fig. 21. The highest value is $J_c = 1.1 \times 10^5$ A·cm$^{-2}$, corresponding to $J_{c,eng} = 1.67 \times 10^4$ A·cm$^{-2}$ at 10 T, where $J_{c,eng}$ is the overall critical density. It follows that MgB$_2$ magnets could reach 10 T at 4.2 K. Based on the present data, it is expected that at 20 K magnets producing fields up to 5 T could also be fabricated. Each one of the three known methods used for the fabrication of industrial MgB$_2$ wires presents individual advantages. *Ex situ* wires show a high homogeneity over long wire lengths and have their best performance at low fields, whereas the advantage of *in situ* and IMD processing consists of an easier introduction of carbon-based additives and thus in higher $J_c$ values at high fields.

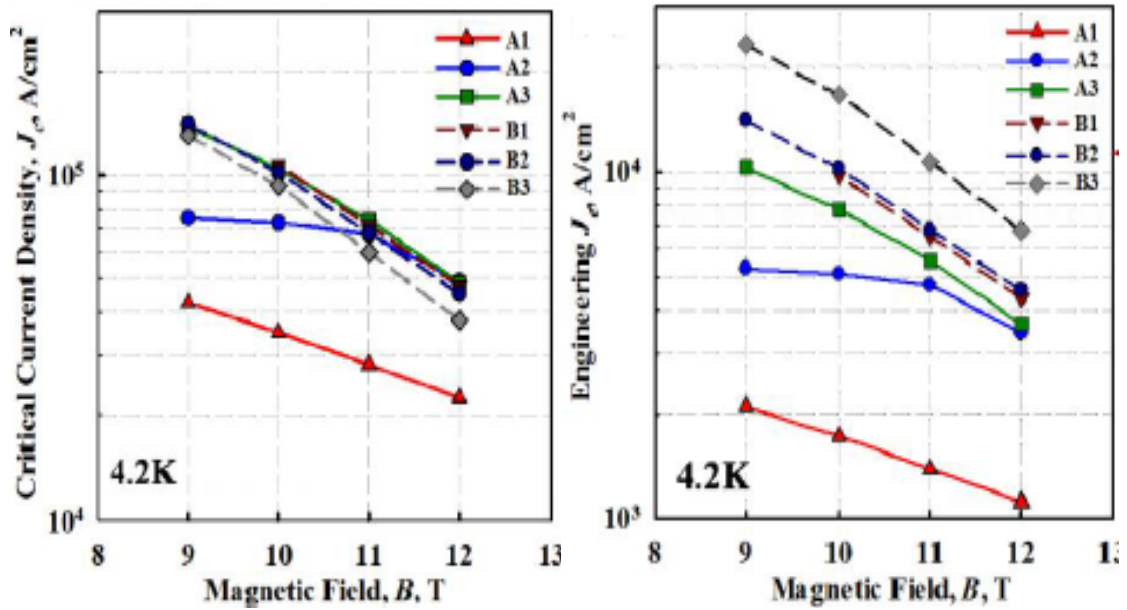

**Fig. 21:** Critical current density of MgB$_2$ prepared by a modified IMD technique. The left figure shows $J_c$ in the filament; the one on the right shows the engineering $J_c$ [19].

The present limits of $J_c$ can be estimated from the data for highly textured thin films: $J_c^{//}$ values for highly textured thin films with $B//ab$ are very similar to those of $Nb_3Sn$; on the contrary, $J_c$ for the same thin films, but measured with the field perpendicular to the *ab* plane, exhibits the lowest values of all. From Fig. 22, it can be estimated that the expected value of $J_c$ for a round, alloyed $MgB_2$ wire at high field, e.g. 16 T, will not exceed $1 \times 10^4$ A·cm$^{-2}$: this indicates that the values of Li *et al.* [19] in Fig. 19 for fully dense IMD wires are very close to the maximum that can be reached with this compound. Possible ways of increasing $J_c$ by artificial pinning are still under study.

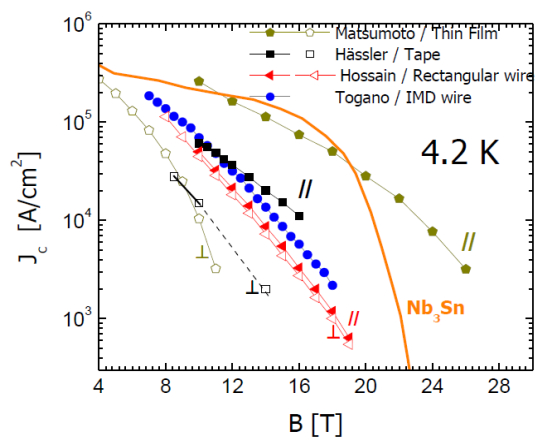

**Fig. 22:** Comparison of $J_c$ vs. $B$ for $MgB_2$ wires and thin films. Thin films: Matsumoto *et al.* [21]; *in situ* tapes: Hässler *et al.* [22]; cold pressed *in situ* wires: Flükiger [18]; IMD wires: Togano [15].

### 3.7 Conclusions for MgB$_2$ wires

In spite of the relatively low value of $B_{c2}$ and the inherent anisotropy of the $MgB_2$ phase with respect to other superconductors, the following advantages of $MgB_2$ wires can be recognized.

Advantages:
- abundant constituents (Mg and B);
- no chemical toxicity;
- low cost with respect to HTS materials (comparable to Nb−Ti);
- applicability at $4.2 \leq T \leq 25$ K;
- persistent mode operation in MRI magnets up to 25 K.

Disadvantages:
- poor thermal stability with respect to those of Nb−Ti or Nb$_3$Sn;
- applicability of $MgB_2$ wires is limited to relatively low fields: ~10 T at 4.2 K and ~4 T at 20 K.

In future, the compound $MgB_2$ may be considered for the following applications:
- level measurement of liquid H$_2$ containers (e.g. hydrogen-powered cars);
- hydrogen-cooled high current leads at $T \approx 20$ K (Russia);
- LINK project (CERN): 13 000 A current leads at ~10 K (total: >1000 km of MgB2 wire)—see the talk by A. Ballarino at the present CAS Summer School [17];
- wind generators; a first generator study for 10 MW is under construction (European project).

At present, $MgB_2$ is a 'niche' superconductor, but the niche is apparently quite large. Going by the projections of Columbus, it is at least 10 times cheaper than HTS conductors. Also, it can be made by the PIT process in round wire or in tape form, thus exhibiting a flexible architecture. And, although it cannot challenge either Nb-Ti or Nb$_3$Sn at 4 K, it can challenge them at 20–30 K. Because of its low cost, it clearly challenges any HTS conductor for low-field applications in this temperature domain too, where MRI and high-current bus bar applications need an affordable and capable superconductor. MgB$_2$ is viable here, where the cost of HTS rules it out for such applications. It is reasonable be optimistic about further significant advances in the application of MgB$_2$ conductors.

## 4   The system Bi-2212

Bi-2212 is the only HTS that allows the fabrication of round wires with isotropic properties. This leads to new possibilities, e.g. it would be possible to build large accelerator magnets at fields well above 15 T, in particular dipoles or quadrupoles. The results presented here have been obtained by D. Larbalestier's group at Florida State University. Additional data can be found in the publications of Malagoli *et al.* [23], Kametani *et al.* [24], Jiang *et al.* [25] and Scheuerlein *et al.* [26].

These works have progressed in collaboration with a manufacturer, Oxford Instruments (OST). All data shown here are extracted from the studies of these working groups. The fabrication process consists of a series of deformation steps of a silver tube filled with previously reacted Bi-2212 powder followed by a final high-temperature heat treatment at $T \leq 888°C$.

After several years of development with relatively deceptive $J_c$ values, progress was achieved after observing that the Bi-2212 filaments post heat treatment were not wire-like, but were rather in the form of hollow 'tubes' which exhibited a great number of holes. Large bubbles form on melting and holding at $T_{max}$, resulting in large holes in the wall of the Bi-2212 'tubes', obviously leading to a strong reduction of $J_c$. The origin of these holes is the formation of a gas phase in this narrow temperature range. A series of high-pressure treatments was tried in order to avoid the formation of holes in the Bi-2212 walls. It was found that a moderate external Ar pressure was sufficient to reach a sizeable increase of $J_c$, as shown in Fig. 25. The progress in Bi-2212 round wires is visualized in Fig. 21, which compares performance at 4.2 K with those of the other known superconductors available in a round shape.

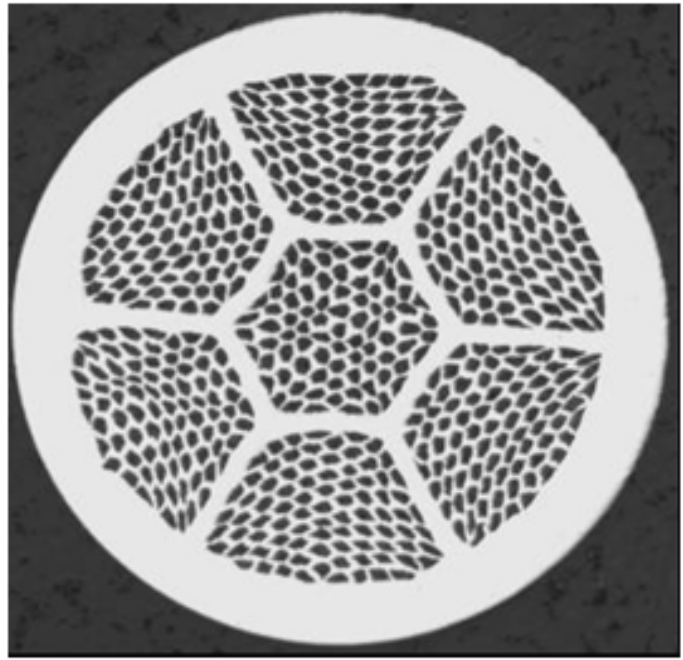

**Fig. 23:** Example of a round multifilamentary Bi-2212 wire, produced by Oxford Instruments OST (USA), before heat treatment, showing the filament configuration. The matrix consists of silver.

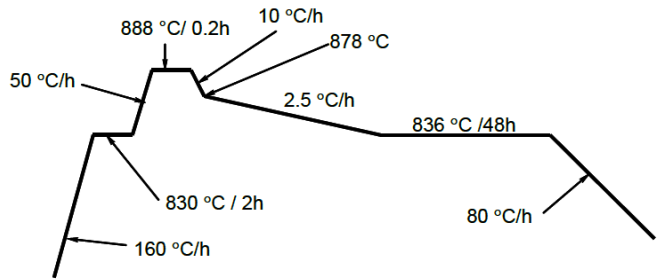

**Fig. 24:** Heat treatment performed on the Bi-2212 wires after the final deformation [23–26]

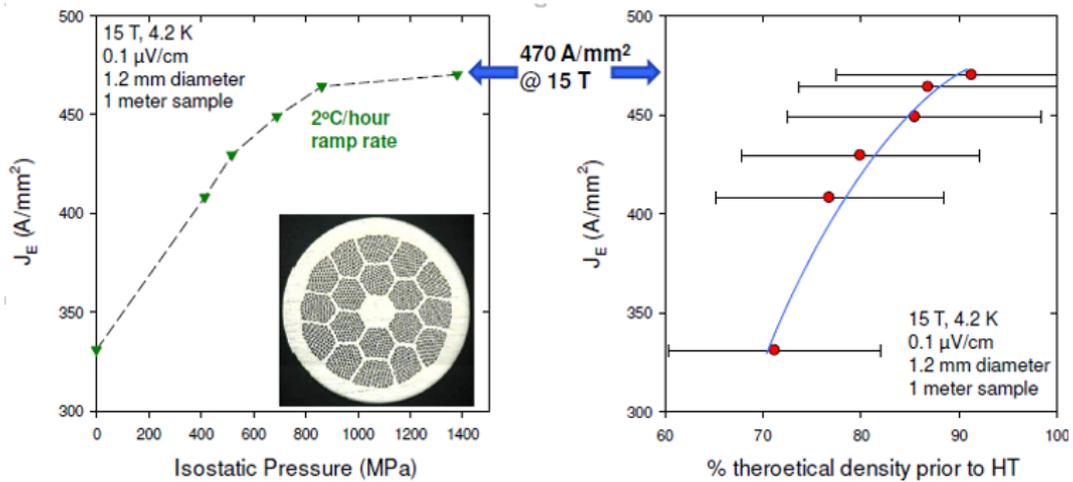

**Fig. 25:** Increase of the engineering critical current density $J_E$ (overall critical current density) of round Bi-2212 wires as a function of the applied isostatic pressure [27].

It can be seen that at 4.2 K, $J_c$ (layer) of Bi-2212 wires is higher than for any of the others for fields exceeding 14 T. For the industrial application, the cross-over with $Nb_3Sn$ wires is at 18 T. Further progress is expected in the future.

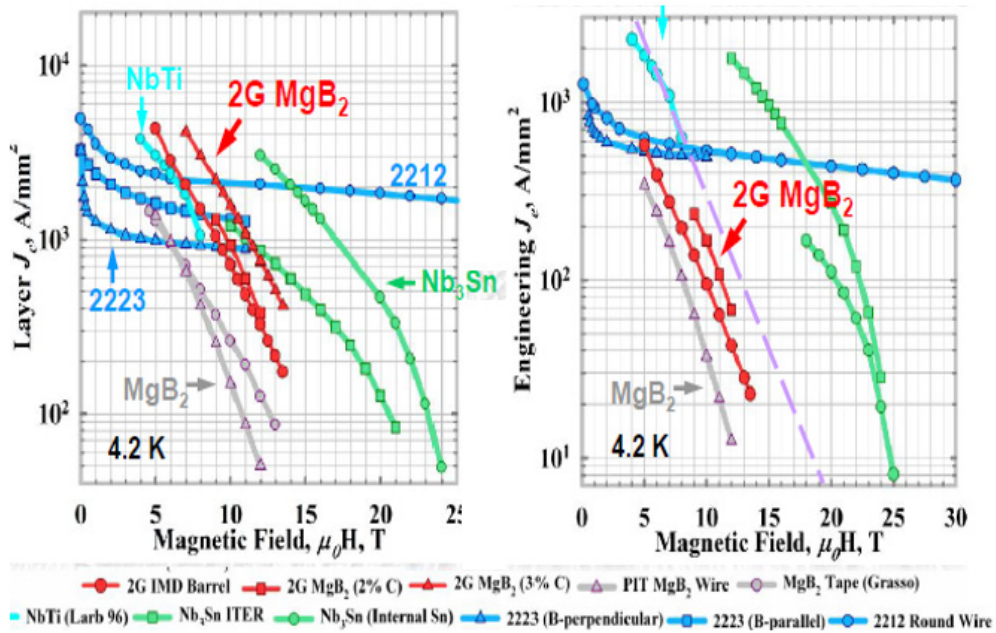

**Fig. 26:** $J_c$ vs. applied field (layer and engineering $J_c$ values) at 4.2 K of various round wires based on known superconducting materials [27].

## 4.1 Conclusions for Bi-2212 wires

On the condition that the recent progress made will lead to the fabrication of homogeneous round wires of kilometre lengths, the following comments can be made [23–27].

Advantages:
- Bi-2212 is the only HTS material available in round wires;
- the multifilamentary configuration is OK;
- excellent thermal stability (Ag sheath);
- at 4.2 K, Bi-2212 exhibits the highest Jc values at B > 18 T of all round shaped superconducting wires;
- after further progress, it may be envisaged that Bi-2212 can be used for fields B > 20 T.

Disadvantages:
- higher costs due to processing and Ag sheath;
- bubble problem: in principle this is solved, but industrial tests on long wire lengths are still required;
- poor mechanical stability: may be the most important obstacle at high-field applications;
- (accelerators): new strategies for mechanical reinforcement must be developed.

## 5 Coated conductor tapes based on R.E.Ba$_2$Cu$_3$O$_y$ superconductors

Magnet applications for high-energy physics are still a major driver for the further development of superconducting technology. High-temperature superconductors (HTS) with very high $B_{c2}$ values are promising with respect to large magnets generating fields exceeding 20 T at 4.2 K, as required in future accelerators following LHC Upgrade. The current-carrying properties of R.E.Ba$_2$Cu$_3$O$_y$ (REBCO) coated conductors with $T_c$ = 92 K are well suited to fulfil these needs. However, these materials cannot yet be obtained as round wires, due to the high anisotropy, which follows directly from the layered crystal structure (see Fig. 27) of the compound YBa$_2$Cu$_3$O$_{7-x}$. (In the applications, the Y in this formula will be replaced by a Rare Earth (R.E.) metal, due to its higher current-carrying properties, so YBa$_2$Cu$_3$O$_{7-x}$ is usually abbreviated to R.E.BCO.)

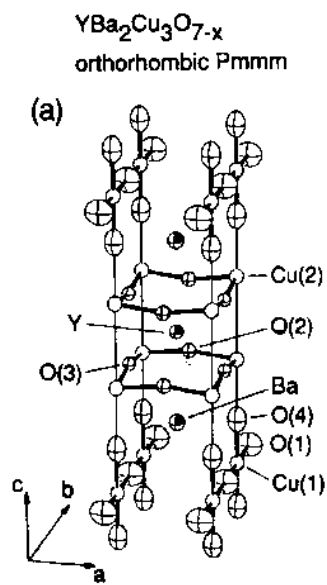

**Fig. 27:** The crystal structure of the HTS compound YBa$_2$Cu$_3$O$_{7-x}$

In contrast to the three systems presented before, R.E.BCO conductors can at present only be prepared in the shape of flat tapes, called coated conductors. It follows that particular techniques must be developed to fabricate dipoles or quadrupoles with R.E.BCO. In the following, the main properties of coated conductors are very briefly reviewed in view of their possible application in accelerators at 4.2 K. A variety of methods have been developed by various manufacturers to fabricate coated conductors; two of them are represented in Fig. 28.

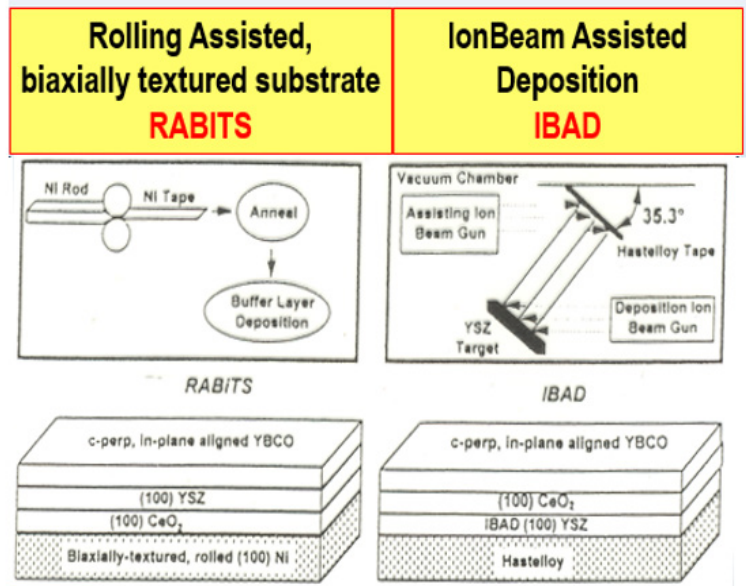

**Fig. 28:** Schematic of RABITs and IBAD fabrication techniques for R.E.BCO (given in the figure as YBCO) coated conductors.

The main difference between the RABITS and the IBAD processes concerns the substrate, which consists of biaxially textured Ni foils and by untextured Hastelloy, respectively. As a consequence, RABITS wires exhibit lower uniaxial tensile stresses than IBAD wires, where irreversible strains for $I_c$ up to 0.7% can be reached. The multilayered architecture of coated conductors fabricated using RABITS and IBAD is shown in Fig. 29; it is easily seen that, due to various problems linked to mechanical stability and to the need for intermediate buffer and stabilizer layers, the total cross-section of the superconductor is limited to values around 1%. In order to enhance the critical current density, it is thus necessary to enhance the HTS layer thickness. However, it is very difficult to maintain a perfect biaxially textured layer for thicknesses exceeding 1 $\mu$m; this has only recently been possible, after very sophisticated and costly improvements of the deposition parameters.

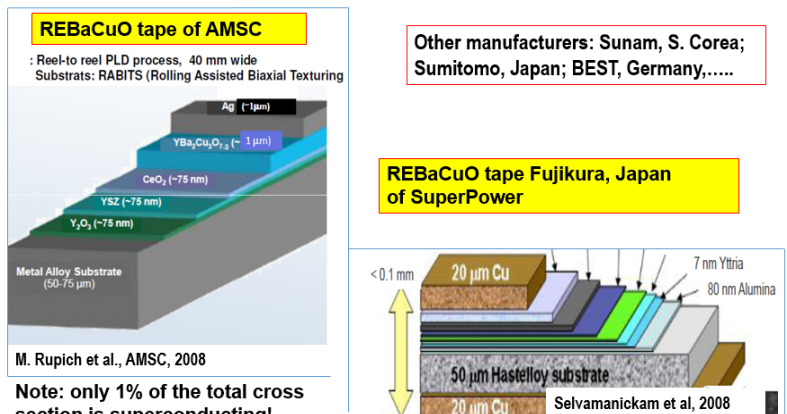

**Fig. 29:** Multilayered architecture of RABITS and IBAD coated conductors. (Thanks to M. Rupich from AMSC and to Selvamanickam from SuperPower.)

It is customary to define the critical current density by the unit $A \cdot cm^{-1}$, where the value is normalized for a 1 cm wide tape ($J_{cw} = J_c \cdot cm^{-2} \times 1\ cm = A \cdot cm^{-1}$), and no longer depends on the tape width. The variation of the critical current density as a function of the HTS layer thickness up to 6 $\mu$m at 77 K is represented in Fig. 30. These values are obtained on short samples; the values on industrial coated conductor lengths (≤500 m) obtained by various manufacturers are presently in the range 500–600 A, with a tendency towards higher values.

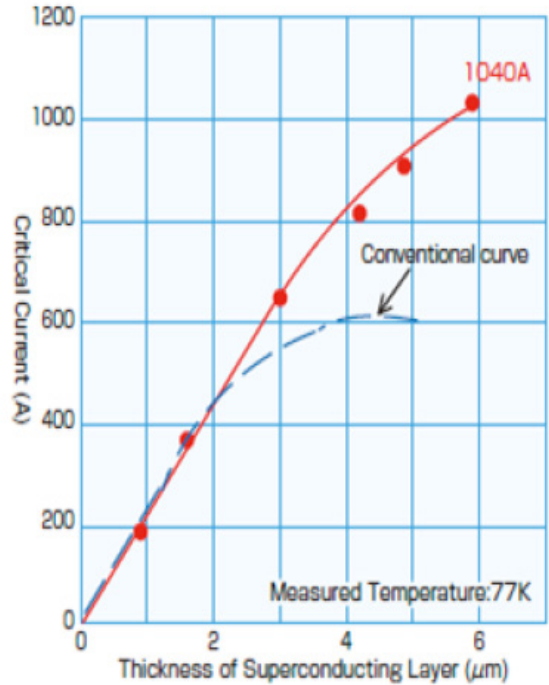

**Fig. 30:** $I_c$ vs. thickness of the HTS superconducting layer at 77 K at zero field. Maximum critical current density obtained $J_c = 1040$ A·(cm-width)$^{-1}$ for a 6 mm thick layer. Definition: $J_{cw} = J_c \cdot cm^{-2} \times 1\ cm = A \cdot cm^{-1}$ is normalized for a tape width of 10 mm. (Courtesy: Fujikura (Japan).)

The variation of $J_c$ for a GdBCO tape of Fujikura (Japan) with an initial value of 600 A·cm$^{-1}$ as a function of applied fields up to 31 T and at various temperatures between 4.2 and 77 K is shown in Fig. 31, illustrating the wide potential of this kind of coated conductor tape.

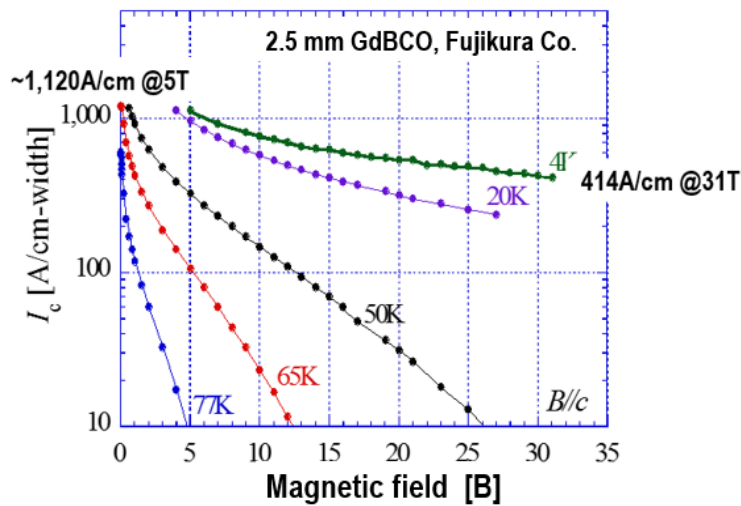

**Fig. 31:** $J_c$ vs. $B$ and $T$ for a HTS coated conductor (Fujikura) with $J_c = 600$ A·cm$^{-1}$. Measurements taken from the group of Professor Kiss (Kyushu University), in collaboration with Florida State University and Tohoku University.

The last property discussed in this manuscript concerns the AC losses in coated conductors. Due to the tape geometry, it is impossible to fabricate conductors with the classical multifilamentary configuration. A common way to lower AC losses is to configure the tapes into a Roebel geometry, shown schematically in Fig. 32. Recently, a step towards a 'multifilamentary' architecture consisted in the so-called 'striation by laser scribing' technique, a carefully adjusted laser process whereby the whole tape width is subdivided into three or more 'filaments'.

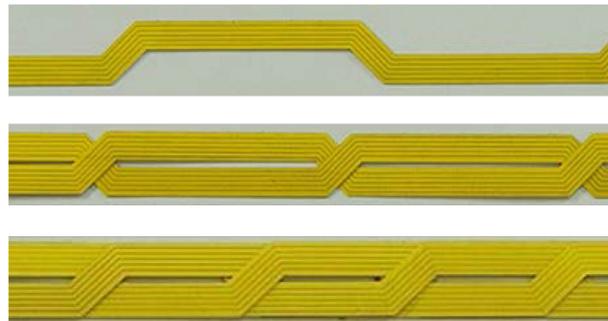

**Fig. 32:** Schematic representation of Roebel + striation for the reduction of AC losses in coated conductors

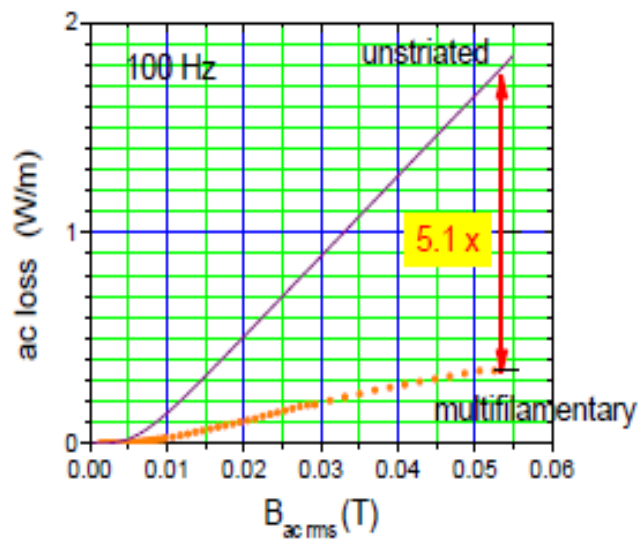

**Fig. 33** Reduction of AC losses by simultaneous Roebel + striation of a coated conductor tape. (Source: Fujikura (Japan).)

The combination of Roebel + striation, also shown in Fig. 32, indeed represents a simulation of a multifilamentary arrangement. It has led to a reduction of the AC losses, as shown in Fig. 33. The reduction of AC losses can be obtained in tapes by means of 6–10 striations. In cables, it is possible to use the Roebel technique in combination with striations.

# 6   Conclusions for coated conductor wires

Due to limited time, only a small number of properties of HTS coated conductors have been discussed in the present lecture. Some conclusions about the future possibilities of these conductors can be drawn. The requirements for HTS coated conductors can be summarized as follows.

*Current density:*
- Improvement with thicker biaxially textured HTS layers.
- Improvement by replacing Y by rare earths (R.E.) (dopants).

- Improvement by introducing artificial pinning centres.

*Mechanical properties:*
- Substrate must be strong enough to withstand the high-temperature formation of REBaCuO.
- The tape as a whole must be strong and flexible enough to be wound into cables and coils at 300 K.
- The tape must withstand longitudinal and transverse stresses during operation.

*Electrical stability*:
- Excess current must be carried in the Ag layer and in the Al, Cu,… outer layers.

*Thermal stability:*
- Heat transfer to the coolant must be optimized.

*AC losses*:
- The architecture must be modified to minimize AC losses (Roebel, striations).

In order to achieve a wide market penetration, a certain number of tasks have to be accomplished by coated conductor tapes:

- higher homogeneity of $J_c$ over whole tape length;
- thicker HTS layer $I_c$ values;
- reproducible production of >1 km lengths;
- enhanced pinning by nano-additives;
- reduced anisotropy by nano-additives.

There is no doubt that all these tasks will be accomplished, taking into account the important efforts undertaken by the industry. However, a major factor monitoring the application volume of coated conductors will be the production costs, which are at present one order of magnitude too high.

**References**

[1] D.M.J. Taylor and D.P. Hampshire, Report No. 8, EFDA03-1126, 2006 (unpublished); available online at http://www.dur.ac.uk/superconductivity.durham/publications.html.
[2] W. Goldacker and R. Flükiger, *Adv. Cryo. Eng*. **28** (1982) 361.
[3] J. Ekin, *Experimental Techniques for Low Temperature Measurements* (Oxford University Press, New York, 2006), pp. 412–467.
[4] D. Uglietti, B. Seeber, V. Abächerli, A. Pollini, D. Eckert and R. Flükiger, *Supercond. Sci. Technol*. **16** (2003)1000.
[5] G. Mondonico, B. Seeber, C. Senatore, R. Flükiger, V. Coarto, G. De Marz and L. Muzzi, *J. Appl. Phys*. **108** (2010) 093906.
[6] D. Uglietti, B. Seeber, V. Abächerli, N. Banno and R. Flükiger, *IEEE Trans. Appl. Supercond*. **15** (2005) 3652.
[7] W. Specking, W. Goldacker and R. Flükiger, *Adv. Cryog. Eng*. **34** (1988) 569.
[8] G. Mondonico, Ph.D. thesis, University of Geneva, 2013.
[9] S.X. Dou, S. Soltanian, J. Horvat, X.L. Wang, S.H. Zhou, M. Ionescu, H.K. Liu, P. Munroe and M. Tomsic, *Appl. Phys. Lett*. **81** (2002) 3419.
[10] E.W. Collings, M.D. Sumption, M. Bhatia, M.A. Susner and S.D. Bohnenstiehl, *Supercond. Sci. Technol*. **21** (2008) 103001.
[11] R. Flükiger and H. Kumakura, in *100 Years of Superconductivity*, Eds. H. Rogalla and P.H. Kes (CRC Press, Taylor and Francis, 2011), p. 758

[12] A. Malagoli, C. Bernini, V. Braccini, C. Fanciulli, G. Romano and M. Vignolo, *Supercond. Sci. Technol.* **22** (2009) 105017.

[13] G. Grasso, A. Malagoli, D. Marrè, E. Bellingeri, V. Braccini, S. Roncallo, N. Scati and A.S. Siri, *Physica C* **378–381** (2002) 899.

[14] M.A. Susner, Y. Yang, M.D. Sumption, E.W. Collings, M.A. Rindfleisch, M.J. Tomsic and J.V. Marzik, *Supercond. Sci. Technol.* **24** (2011) 012001.

[15] K. Togano, J.M. Hur, A. Matsumoto and H. Kumakura, *Supercond. Sci. Technol.* **22** (2009) 015003.

[16] G. Giunchi, S. Ceresara, G. Ripamonti, A. DiZenobio, S. Rossi, S. Chiarelli, M. Spadoni, R. Wesche and P.L. Bruzzone, *Supercond. Sci. Technol.* **16** (2003) 285.

[17] A. Ballarino, LINK project, presentation at CAS 2013.

[18] R. Flükiger, M.S.A. Hossain, C. Senatore and M. Rindfleisch, *Physica C* **471** (2011) 222011.

[19] M.S.A. Hossain, C. Senatore, M. Rindfleisch and R. Flükiger, *Supercond. Sci. Technol.* **24** (2011) 075013.

[20] G.Z. Li, M.D. Sumption, M.A. Susner, Y. Yang, M. Kongara, M.A. Rindfleisch, M.J. Tomsic, C.J. Thong and E.W. Collings, *Supercond. Sci. Technol.* **26** (2012) 115023.

[21] A. Matsumoto, Y. Kobayashi, K.-I. Takahashi, H. Kumakura and H. Kitaguchi, *Appl. Phys. Express* **1** (2008) 021702.

[22] W. Hässler, P. Kovac, M. Eisterer, A.B. Abrahamsen, M. Herrmann, C. Rodig, K. Nenkov, B. Holzapfel, T. Melisek, M. Kulich, M. v. Zimmermann, J. Bednarcik and J.C. Grivel, *Supercond. Sci. Technol.* **23** (2010) 065011.

[23] A. Malagoli, C. Bernini, V. Braccini, G. Romano, M. Putti, X. Chaud and F. Debray, *Supercond. Sci. Technol.* **24** (2011) 075016.

[24] F. Kametani, T. Shen, J. Jiang, C. Scheuerlein, A. Malagoli, M. Di Michiel, Y. Huang, H. Miao, J.A. Parrell, E.E. Hellstrom and D.C. Larbalestier, *Supercond. Sci. Technol.* **24** (2011) 075009.

[25] J. Jiang, W.L. Starch, M. Hannion, F. Kametani, U.P. Trociewitz, E.E. Hellstrom and D.C Larbalestier, *Supercond. Sci. Technol.* **24** (2011) 082001.

[26] C. Scheuerlein, M. Di Michiel, M. Scheel, J. Jiang, F. Kametani, A. Malagoli, E.E. Hellstrom and D.C. Larbalestier, *Supercond. Sci. Technol.* **24** (2011) 115004.

[27] D. Larbalestier, oral presentation at CERN, November 2012.